\documentclass[usenatbib]{mn2e} 
\input psfig.sty 
\usepackage{epsfig}
\usepackage{amsmath,amssymb} \usepackage{psfig}

\newcommand{\kpc} {{\,\rm kpc}} 
\newcommand{\pc} {{\,\rm pc}} 
\newcommand{\kms}{{\,\rm {km\,s^{-1}} }} 

\title[Cold streams and clumpy galaxies] {Disk formation and the origin of clumpy galaxies at high redshift}\author[Oscar Agertz
  et al.]  {\parbox[t]{\textwidth}{Oscar
Agertz$^1$\thanks{agertz@physik.unizh.ch}, Romain Teyssier$^{1,2}$, Ben Moore$^1$}\vspace*{6pt}\\$^1$ Institute for Theoretical Physics, University of Z\"urich, CH-8057 Z\"urich, Switzerland\\$^{2}$ CEA Saclay, DSM/IRFU/SAp, Batiment 709, 91191 Gif-sur-Yvette Cedex, France}
\date{\today}
\begin{document}
\maketitle
\begin{abstract} 
Observations of high redshift galaxies have revealed a multitude of large clumpy rapidly star-forming galaxies. Their formation scenario and their link to present day spirals is still unknown. In this \emph{Letter} we perform adaptive mesh refinement simulations of disk formation in a cosmological context that are unrivalled in terms of mass and spatial resolution. We find that the so called `chain-galaxies' and `clump-clusters' are a natural outcome of early epochs of enhanced gas accretion from cold dense streams as well as tidally and ram-pressured stripped material from minor mergers and satellites. Through interaction with the hot halo gas, this freshly accreted cold gas settles into a large disk-like system, not necessarily aligned to an older stellar component, that undergoes fragmentation and subsequent star formation, forming large clumps in the mass range $10^7-10^9\,M_\odot$. Galaxy formation is a complex process at this important epoch when most of the central baryons are being acquired through a range of different mechanisms - we highlight that a rapid mass loading epoch is required to fuel the fragmentation taking place in the massive arms in the outskirts of extended disks, an accretion mode that occurs naturally in the hierarchical assembly process at early epochs.
\end{abstract}


\section{Introduction}
\label{sect:intro}
The morphology and star formation properties of high redshift galaxies are very different from present day quiescent spirals and ellipticals. Large clumpy irregular disks with kpc-sized star forming clumps as massive as $M_{\rm cl}\sim10^7-10^9\,M_\odot$ are observed in the Hubble Ultra Deep Field (UDF) \citep[e.g.][]{DebraElmegreen07,Elmegreen09}, a population that is very rare today.
`Chain galaxies', first identified by \cite{cowie95}, are believed to be high-redshift disky galaxies seen edge-on, while `clump cluster' galaxies are their face on counterparts \citep{dalcanton96,Elmegreen04b}. In optically selected samples, high redshift galaxies show very high star formation rates up to $100-200\,M_\odot{\rm yr}^{-1}$ \citep{Daddi04} and in recent spectroscopic observations they appear to be extended, though perturbed, rotating disks \citep{Forster06,Genzel06,Genzel08}. The origin of these galaxies and how they connect and possibly evolve into present day spirals is still unknown. Gas rich major mergers give rise to large, bulge-dominated rotating disks \citep[]{Robertson08} even though massive clumps can form at large radii, from globular clusters \citep{Bournaud08b} to tidal dwarf galaxies \citep{Elmegreen93,Barnes92}. However, major mergers are not frequent enough \citep{Dekel09} and are more likely to be the origin of the rare, extremely high star forming, sub-millimeter galaxies \citep{Zheng04,Jogee08}.

Observational evidence \citep{Elmegreen06,Bournaud08,Shapiro08} suggests that clumps form in gas rich spiral disks rather than during on-going mergers, although the latter scenario can not be completely ruled out \citep{Taniguchi01,Overzier08}. Recent work by \cite{Bournaud07} (hereafter B07) and \cite{Elmegreen08} has demonstrated that internal disk fragmentation can effectively reproduce many of the observables of chain and clump clusters galaxies and that these different clumpy systems can have the same origin but observed at different inclinations. However, the models of B07 still rely on idealized, pre-existing very massive gas disks, in order to reproduce the massive clumps and can not explain an ongoing, steady-state fragmentation scenario.

How galaxies acquire their baryons is an open question. The classic picture of galaxy formation within the cold dark matter (CDM) scenario assumes that the accreted gas is shock heated to the virial temperature, cools radiatively and rains down to form an inner star-forming rotating disk. Recent theoretical studies \citep{BirnboimDekel03,Keres05,DekelBirnboim06,Ocvirk08,Keres09,Brooks08,Dekel09} have demonstrated that accretion of fresh gas via cold infall can in fact be the dominant process for gas accretion for halo masses $M\lesssim10^{11.6}\,M_\odot$. In these halos, the cooling time for gas of temperature $T\sim10^{4}\,{\rm K}$ is shorter than the timescale of gas compression and shocks are unable to develop. In halos above this mass, cold accretion persists as gas is supplied by cold streams penetrating through hot massive halos at $z\gtrsim2$ \citep{Ocvirk08,Dekel09} whilst the classical hot mode of gas accretion dominates at lower $z$. Because of insufficient spatial resolution, these studies could not follow the evolution of the accreting gas and how the cold streams connect to the central galaxies. The purpose of this \emph{Letter} is to look in detail at the gas accretion and disk formation process using state-of-the-art numerical simulations. 
\label{sect:results}
\begin{figure*}
\center
\psfig{file=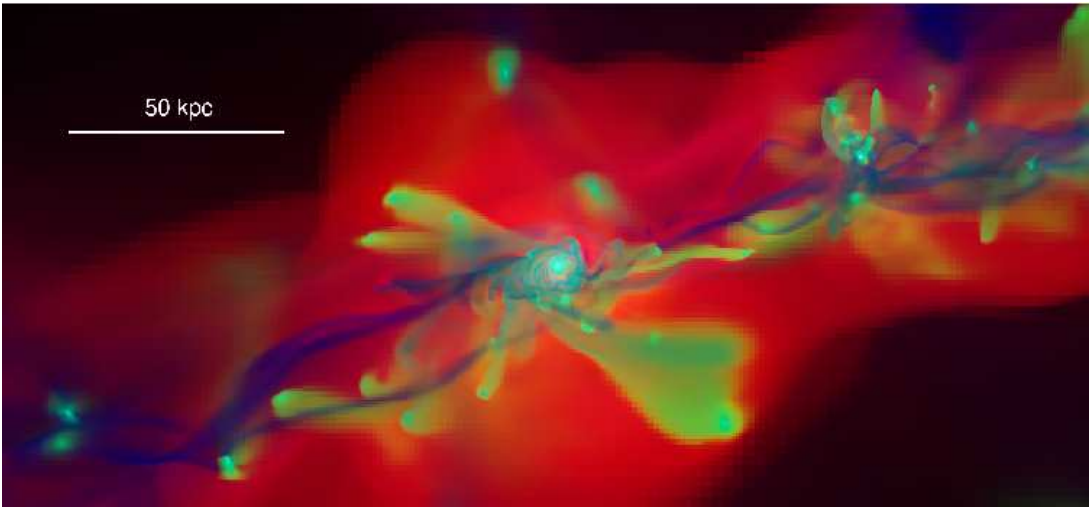,width=0.82\linewidth} 
\caption[]{An \emph{RGB}-image of the gas showing the disk and accretion region at $z\sim3$. The image is constructed using \emph{R}=temperature, \emph{G}=metals and \emph{B}=density. We can clearly distinguish the cold pristine gas streams in blue connecting directly onto the edge of the disk, the shock heated gas in red surrounding the disk and metal rich gas in green being stripped from smaller galaxies interacting with the halo and streams of gas. The disk and the interacting satellites stand out since they are cold, dense and metal rich.}
\label{fig:streams}
\end{figure*}
\section{Numerical simulation}
\label{sect:sim}
We use the adaptive mesh refinement (AMR) code {\tt RAMSES} \citep{teyssier02} to simulate the formation of a massive disk galaxy in a cosmological context including dark matter, gas and stars. The gas dynamics are calculated using a second-order unsplit Godunov method, while collisionless particles (including stars) are evolved using the Particle-Mesh technique. The modelling includes realistic recipes for star formation \citep{Rasera06}, supernova feedback and enrichment \citep{Dubois08}. Metals are advected as a passive scalar and are incorporated self-consistently in the cooling and heating routine, as in \cite{Agertz09}, and we adopt an initial metallicity of $Z=10^{-3}Z_\odot$ in the high-resolution region. The refinement strategy is based on a quasi-Lagrangian approach, so that the number of particles per cell remains roughly constant, avoiding discreteness effects \citep[e.g.][]{Romeo08}. The computational domain is a 40 Mpc cube containing nested AMR grids of particles and gas cells down to a Lagrangian region containing dark matter particles of mass  $m_{\rm p}=2.2\times 10^5\,M_\odot$. The effective resolution of our initial grid is therefore $2048^3$. We then refine this base grid according to our refinement strategy, so that the maximum resolution is $\Delta x\sim 40\,\pc$ in {\it physical units} at all times.

For our initial conditions we take the Via-Lactea II simulation \citep{DiemandNature08} which forms a Milky Way sized dark matter halo that accretes most of its mass ($M_{\rm vir}=2\times 10^{12}\,M_\odot$ at $z=0$) by redshift $z=2$. We evolved the entire simulation to $z=0$ at a coarser resolution, here we report on the high redshift evolution to $z=2$ at which point it hosts a disk that is massive enough to be compared to the observations in e.g. \cite{Bournaud08} and \cite{Genzel06}. We use standard galaxy formation ingredients, with a star formation efficiency of 2\% \citep[as defined in][]{Rasera06}, a star formation density threshold $n_{\rm H} = 4\,{\rm cm}^{-3}$ and a supernovae mass loading factor $f_w=10$ \citep[as defined in][]{Dubois08}. In order to prevent artificial fragmentation, we use a pressure floor $P\simeq 3G\Delta x^2 \rho^2$, so that we satisfy the \cite{truelove97} criterion at all times.

\section{Results}
Fig.\,\ref{fig:streams} shows a large scale view of the galactic disk at $z\sim3$. At this time the dark matter halo has reached a mass of $M_{\rm vir}\sim3.5\times10^{11}\,M_\odot$, while the total baryonic mass in the disk (disregarding the bulge) is $M_{\rm bar}\sim2.4\times10^{10}\,M_\odot$ out of which $50\%$ is gas, putting it in a regime where both cold flows and stable shocks can exist \citep{Keres05,DekelBirnboim06,Ocvirk08}. This striking image ties together many aspects present in modern theories of galaxy formation and highlights new complexities. Cold streams of gas originating in narrow dark matter filaments, effectively penetrate the halo and transport cold metal-poor gas right down to the proto-galactic disk to fuel the star forming region.  A comparable amount of metal enriched material reaches the disk in a process that has previously been unresolved - material that is hydrodynamically stripped from accreting satellites, themselves small disky systems, through the interaction with the hot halo and frequent crossings of the cold streams.

Streams of cold gas flow into the halo on radial trajectories, eventually forming orderly rotational motion in an extended disk. This gas is in approximate pressure equilibrium with the hot halo that
has a rotational velocity of $v_{\rm rot}\sim 30\kms$ close to the virial radius, increasing smoothly to $v_{\rm rot}\sim200\kms$ to match the rotation at the edge of the disk ($r\sim10\kpc$). The ram pressure is significant close to the disk, forcing the streams to curve around it. At early times, when the interaction region close to the disk is tenuous, streams can `swing' past the proto-disk before
being decelerated completely. At later times the turbulent accretion region carries significant mass and infalling cold gas quickly decelerates by plowing through it. We detect compression and radiative shocks that quickly dissipate since the cooling times are very short, resulting in a denser configuration for the cold gas. The global outcome of these interaction is a turbulent gas heavy disk prone to fragmentation.
\begin{figure}
\center
\begin{tabular}{cc}
\psfig{file=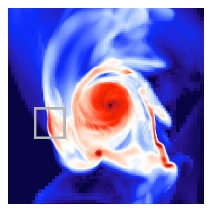,width=85pt} &
\psfig{file=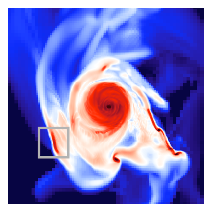,width=85pt} \\
\psfig{file=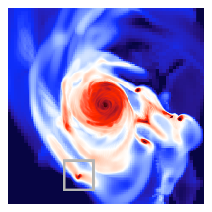,width=85pt} &
\psfig{file=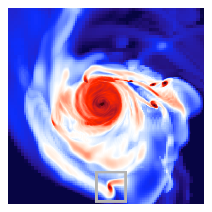,width=85pt} \\
\end{tabular}
\caption[]{A time sequence spanning 40 Myr of the projected gas density at $z\sim 3$ in a $18\times 18\,\kpc^2$ region. The box shows the formation of a $\sim 10^8\,M_\odot$ clump via gravitational instability.}
\label{fig:mapsfrag}
\end{figure}
Fig.~\ref{fig:mapsfrag} shows a time sequence of the complicated and asymmetric gas flows around the gas disk at $z\sim 3$. The figure reveals that many of the large scale spiral arms at large radii are not waves, but material arms that can survive for an orbital time and that these arms are gravitationally unstable and can fragment into clumps. Gravitational instability has been used by \cite{Elmegreen93} (hereafter E93) to explain the formation of massive clumps, as large as dwarf galaxies, in the tidal tails of merging galaxies. The typical mass of objects that form within the arms is $M_{\rm J} \simeq \sigma_{\rm eff}^4/G^2\Sigma$, where $\Sigma$ is the surface density of gas within the arm and the effective mass-weighted 1D velocity dispersion is defined as $\sigma^2_{\rm eff}=c^2_{\rm s}+\sigma^2_{\rm 1D}$ where $c_{\rm s}$ is the local sound speed. Using the small region highlighted by the grey square in Fig.~\ref{fig:mapsfrag}, we have measured $\Sigma = 60\,M_\odot$ pc$^{-2}$ and $\sigma_{\rm eff} \simeq 25\kms$, giving $M_{\rm J} \simeq 2\times10^8\,M_\odot$ which agrees well with the mass of the forming clump. 
The internal dispersion velocity is roughly equal to the divergent motions across the curved gas filament. The typical velocity dispersion across a $\lambda\sim1$ kpc patch of the filament will be of the order $\sigma \simeq \lambda v_{\rm orb}/{\cal R}_{\rm c}\simeq 20\,\kms$, where the orbital velocity $v_{\rm orb}\sim200\,\kms$ and the curvature radius ${\cal R}_c$ of the filament equals the radius of the extended disk. This value agrees well with the dominating turbulent component of $\sigma_{\rm eff}$. As the interaction region grows in mass and develops a more symmetric disk-like morphology we also observe massive clump formation in fragmenting spiral waves at intermediate radii. 
\begin{figure}
\centering
\begin{tabular}{cc}
\psfig{file=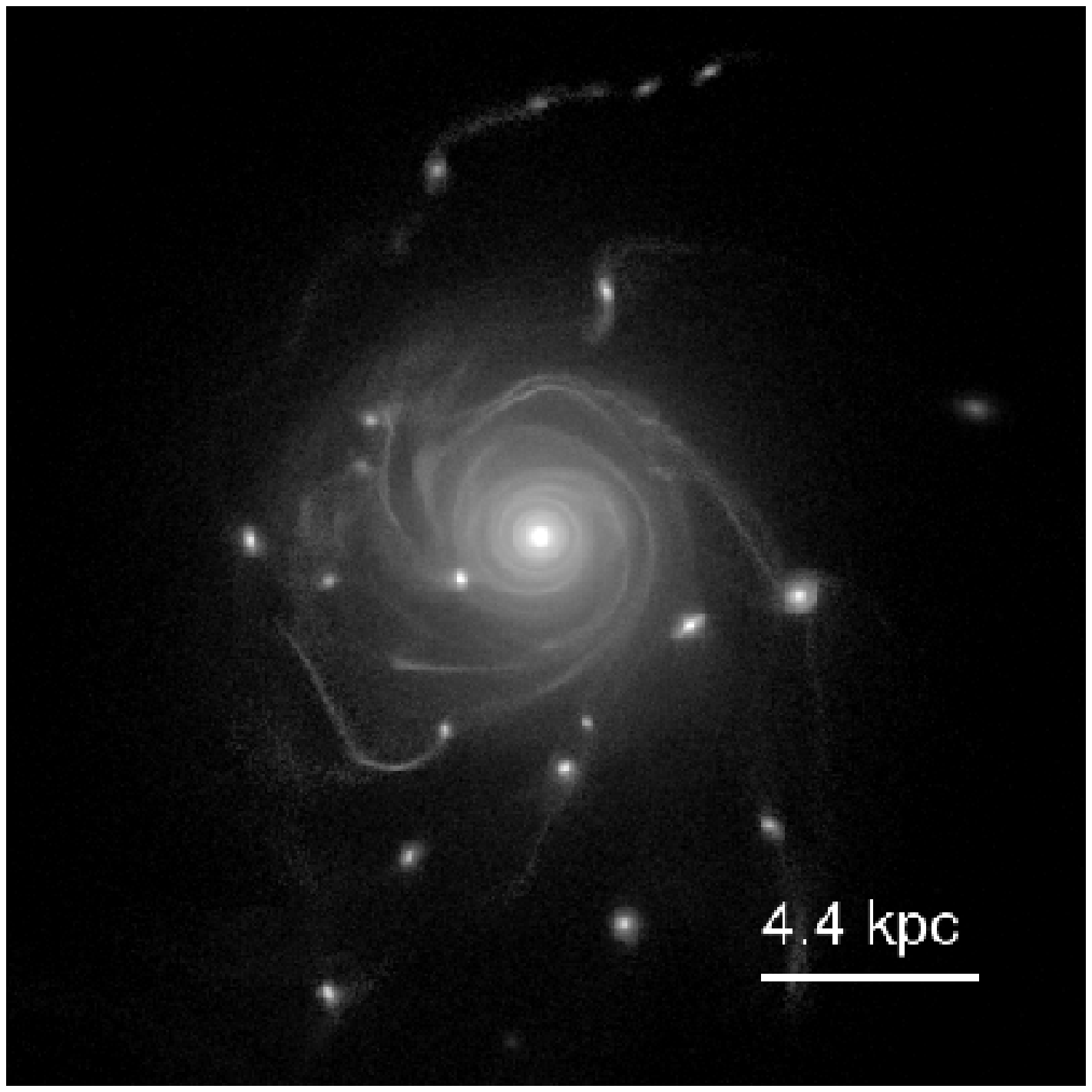,width=85pt} &
\psfig{file=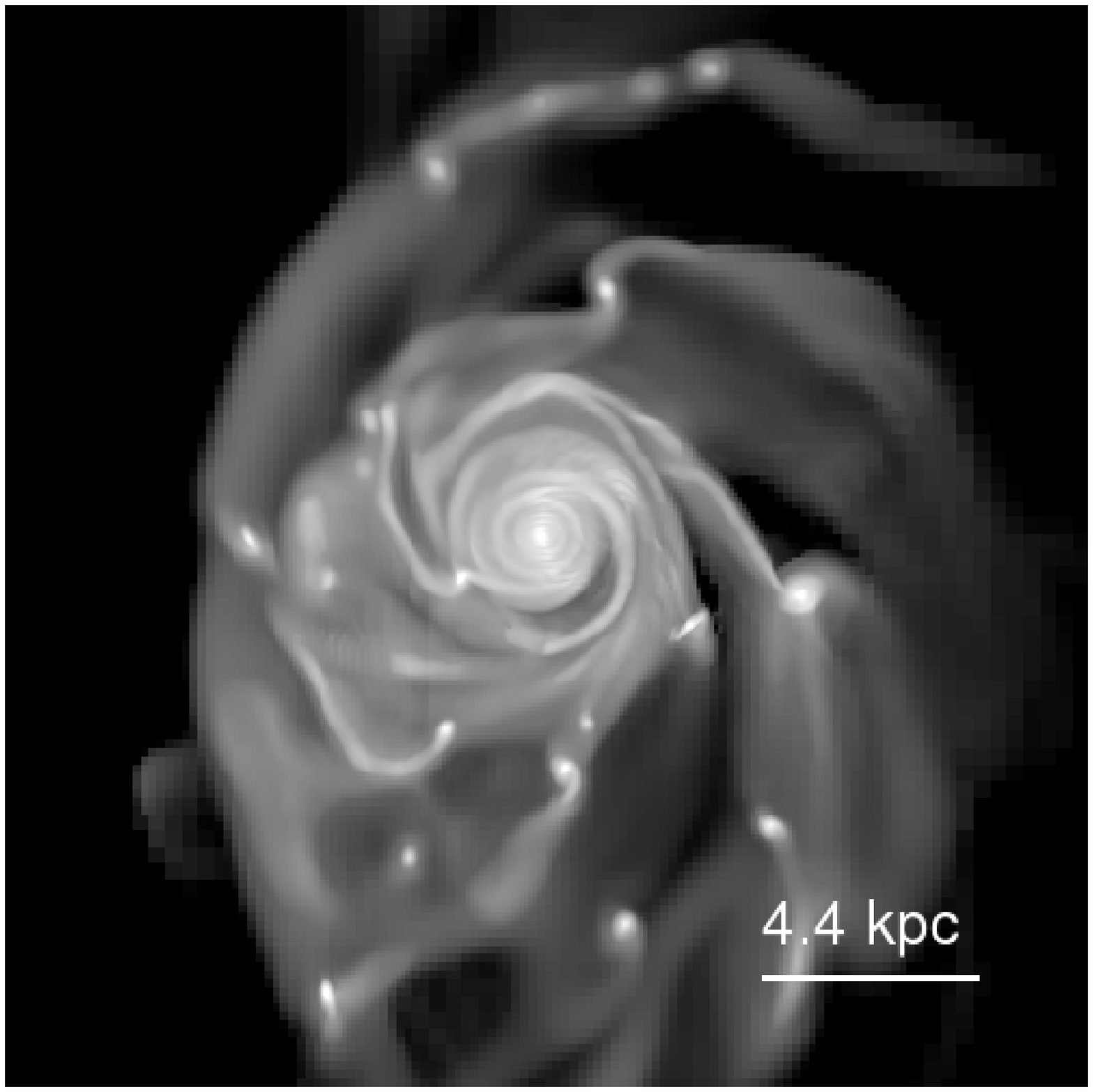,width=85pt} \\
\psfig{file=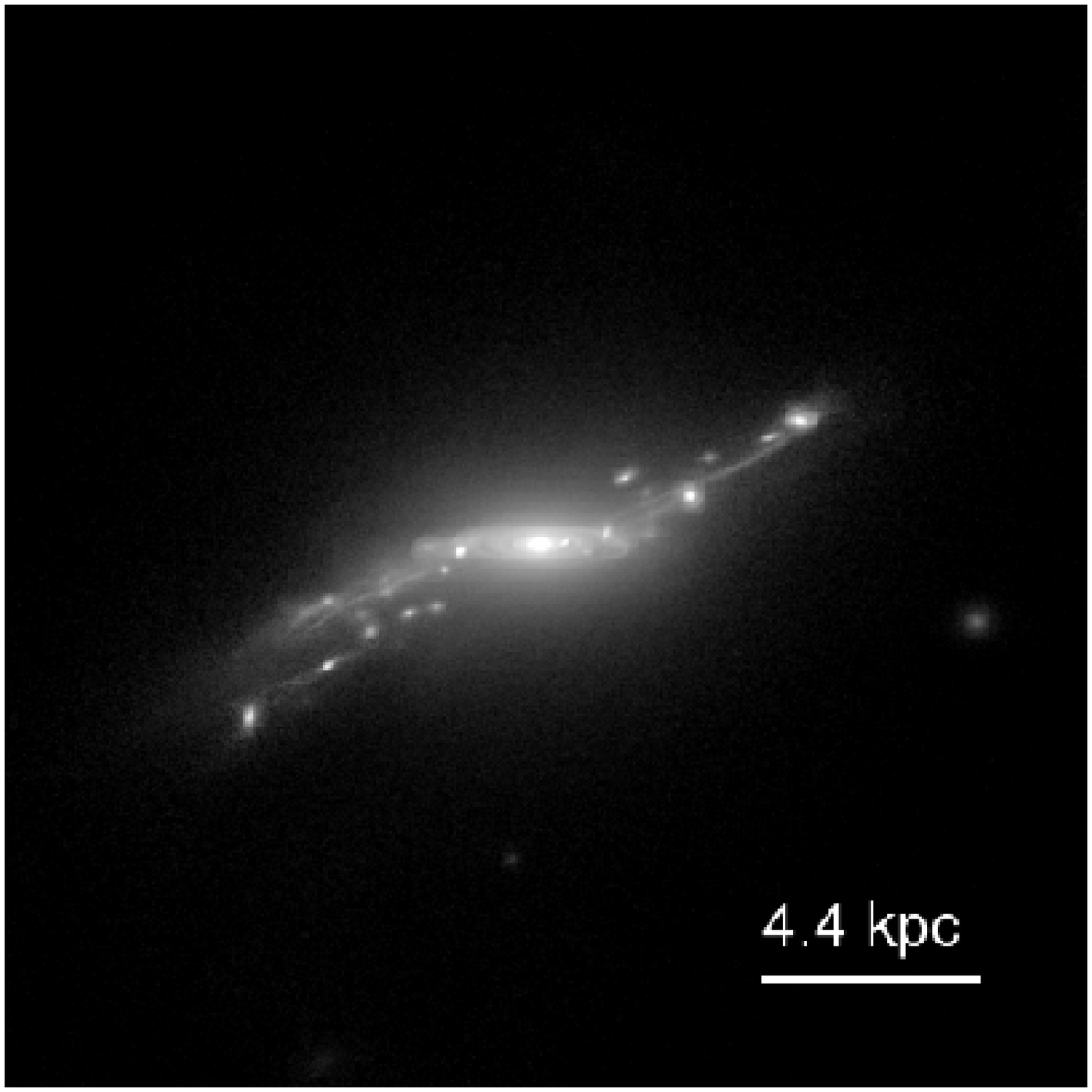,width=85pt} &
\psfig{file=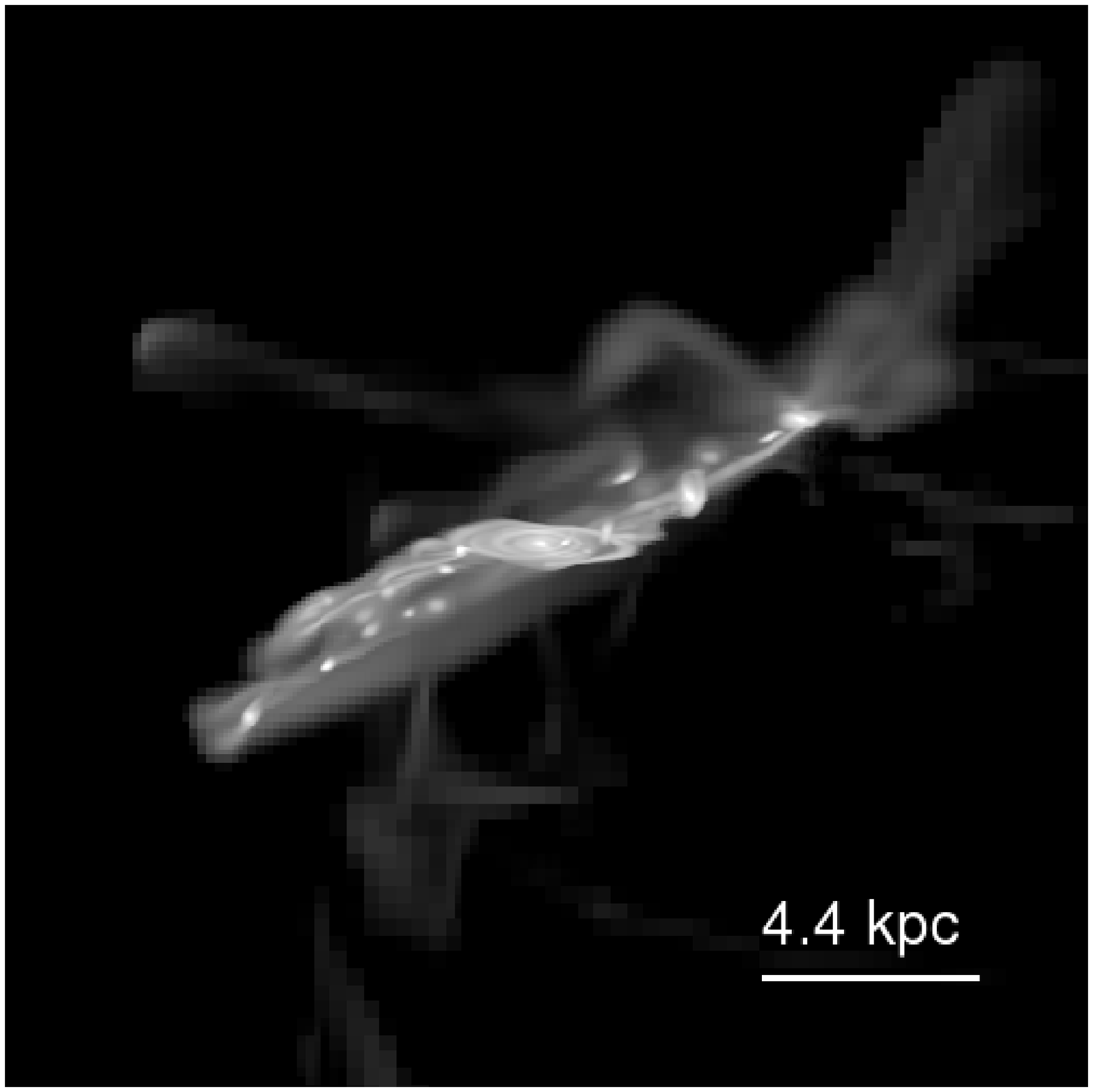,width=85pt} 
\end{tabular}
\caption[]{Density projection of the stars (left-hand panels) and gas (right-hand panels) at $z\sim 2.7$ illustrating the fragmentation process and the formation of large clumps of mass $\sim 10^7-10^9\,M_\odot$. 
}
\label{fig:mapschain}
\end{figure}
The resulting galaxy is shown in Fig.~\ref{fig:mapschain} at $z\sim2.7$, after many large clumps have formed through the above mechanisms. We detect 14 clumps with masses between $M_{\rm cl}\sim5\times10^7$ and $10^9M_\odot$, of which only the two smallest did not form \emph{in situ} but were infalling satellites. The three most massive clumps have $M_{\rm cl}\sim7-8\times10^8\,M_\odot$. In total $\sim15\,\%$ of the baryons are in clumps. In the interaction region between the disk and the cold streams, the typical arm surface density and velocity dispersion can both be estimated using mass average quantities within cylindrical shells. At ($r\sim 8\,\kpc$) we measure $\left<\Sigma\right> \simeq 20$ M$_\odot$ pc$^{-2}$ and $\left<\sigma\right> \simeq 30$ km s$^{-1}$, giving rise to clump masses as large as $M_{\rm J} \simeq 10^9\,M_\odot$. Even though we satisfy the Truelove criterium, convergence in 
the details of the clump properties can be influenced by numerical fragmentation and may require more cells per Jeans length. In addition, numerical diffusion from bulk flows can lead to an underestimation of the turbulent velocity dispersion. Quantifying this is beyond the scope of this paper. The detected clumps are located in the interaction region between the inner disk and the cold streams. In our case this region is not aligned with the initial galactic disk, giving rise to a misalignment of the clumps with respect to the inner galactic disk (see edge-on images in Fig.~\ref{fig:mapschain}). Although we believe that this misalignment is not typical, it is an elegant explanation for the formation of `bent' chains, such as the one reported in \cite{Bournaud08}. Indeed, \cite{Elmegreen06} report that the typical chain galaxy has clumps mostly aligned in the midplane, while in some cases, clumps are seen above and below the midplane (outer and inner disk misaligned). In our case, the misalignment is due to a third cold stream that is perpendicular to the main filament seen in Fig.~\ref{fig:streams}. In a similar scenario, this process has also been invoked to explain the formation of large polar rings \citep{Maccio06}.

The simulated galaxy is sharing many properties with observed chain and clump cluster galaxies \citep{DebraElmegreen07}. Viewed edge-on, the misaligned disk morphology is clearly seen and the overall structure resembles a large chain-galaxy. Viewed face on the spiral-like structure has a similar morphology as clump clusters or clumpy spirals. \cite{Elmegreen05} report that UDF clumpy galaxies at $z\sim1.5-3$ have a stellar mass $\simeq 6\times 10^{10}\,M_\odot $ and a radius $\sim 10\,\kpc$, in striking agreement with our simulated galaxy. Not only does the cosmological simulation reproduce the observed clumpy morphology and global rotation of these systems but we also find a realistic metallicity gradient and star formation rate of 20 $M_\odot {\rm yr}^{-1}$. The inner disk has on average solar metallicity, while that in the clump forming region is only $\sim 1/10\,Z_\odot$, due to the accretion of pristine gas in the cold streams mixing with stripped satellite gas. This has the important observational consequence that these massive clumps might be devoid of dust, making them easier to detect. 

\begin{figure*}
\begin{tabular}{ccc}
\psfig{file=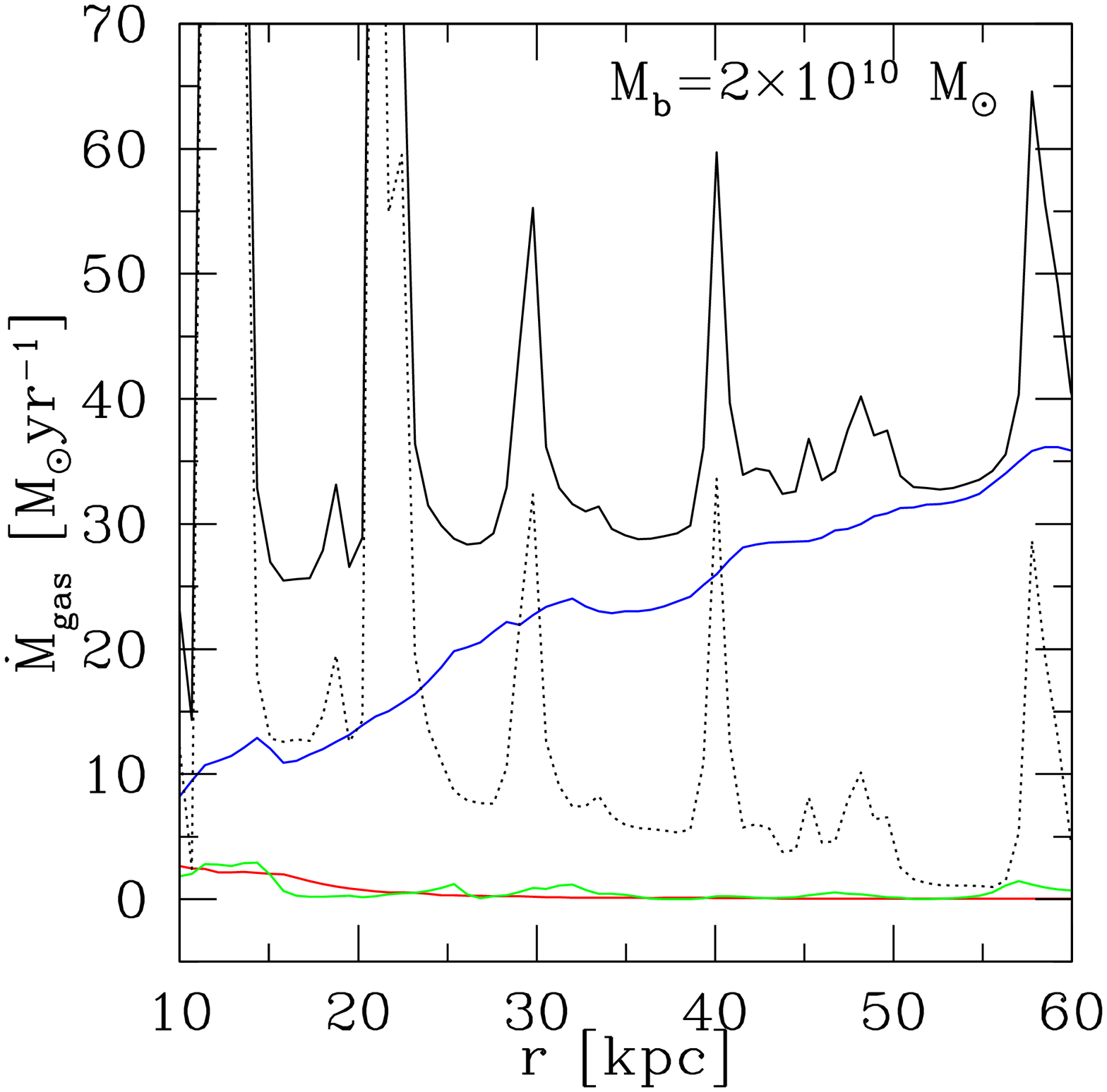,width=145pt} &
\psfig{file=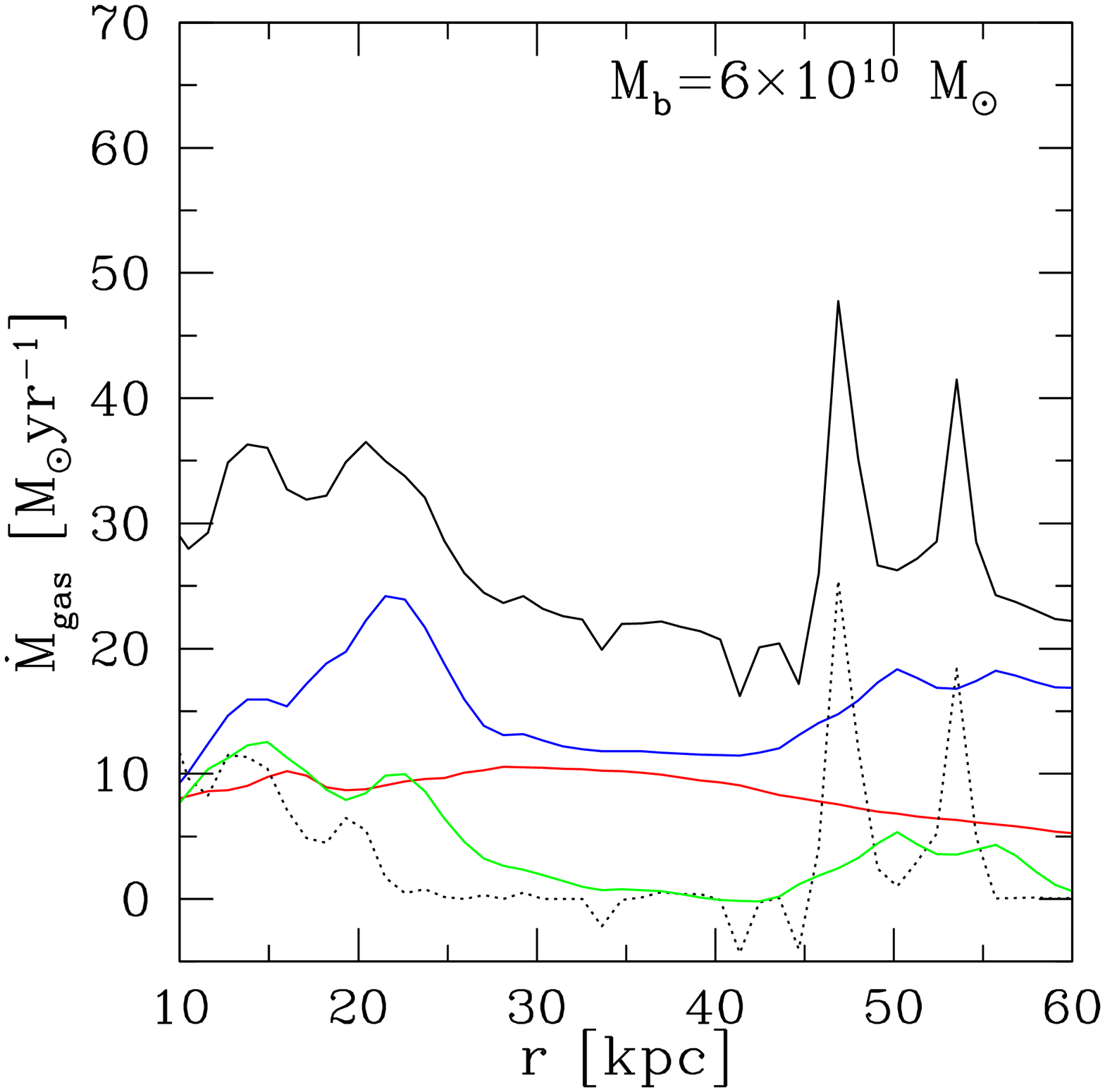,width=145pt} &
\psfig{file=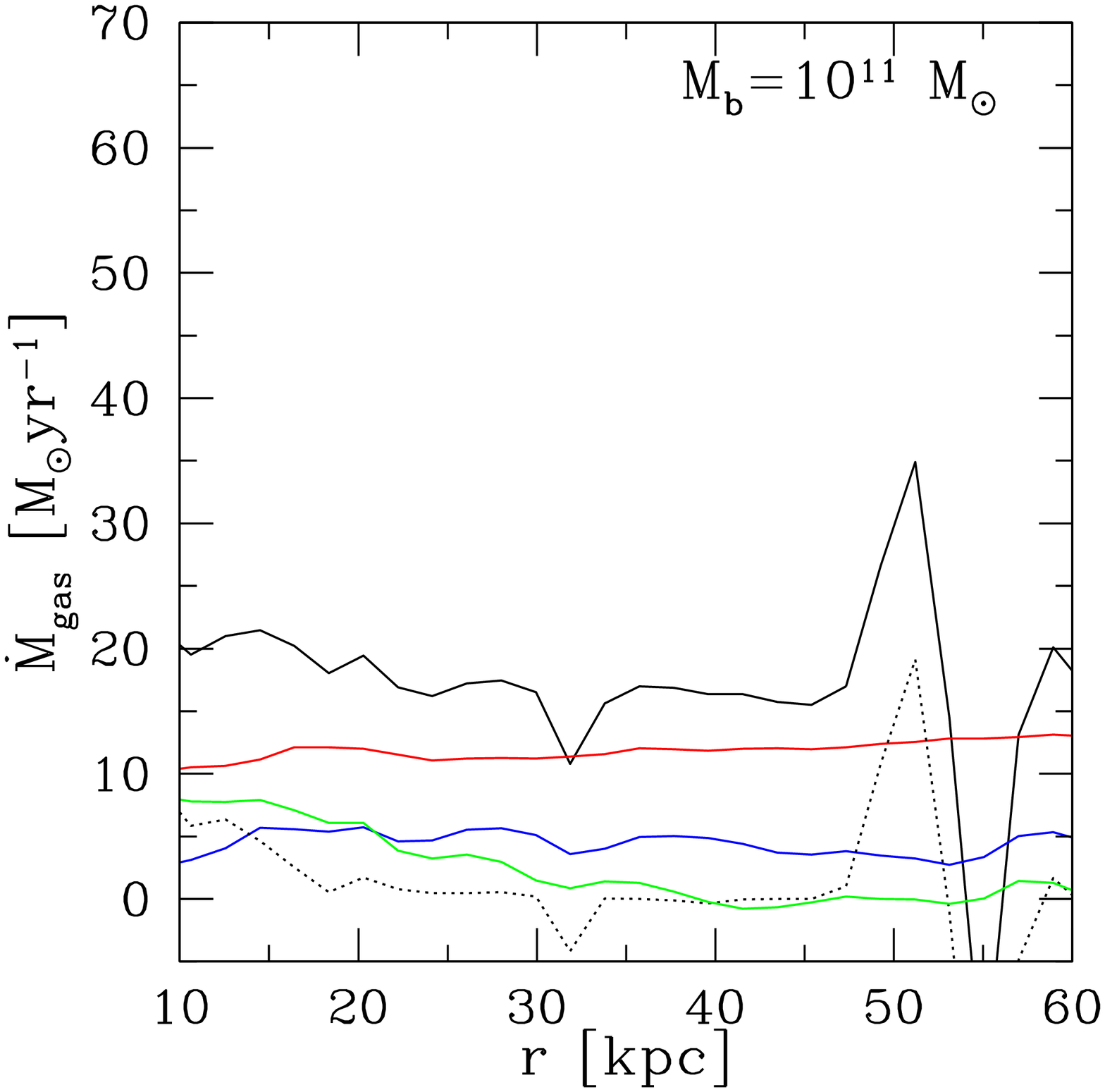,width=145pt}  \\
\end{tabular}
\caption[]{Mass accretion averaged within spherical shells at redshifts $z\sim 5,3$ and 2. The radii are in physical kpc. The lines show the total mass flow (solid black) in each shell, cold diffuse (blue solid), hot diffuse (red solid), dense (dotted) and stripped gas (green) (see text for definition). We observe a decrease in the overall inflow of material and a change from 
cold to hot accretion over time.}
\label{fig:flux}
\end{figure*}

To illustrate `how disks acquire their baryons', we have plotted the mass accretion rate in different gas phases measured around our simulated galaxy at $z=5, 3$ and 2 in Fig.\,\ref{fig:flux}. We define the phases as cold diffuse ($T<2\times10^5\,$K, $n<0.05\,{\rm cm}^{-3}$), dense ($n>0.05\,{\rm cm}^{-3}$), hot diffuse ($T>2\times10^5\,$K, $n<0.05\,{\rm cm}^{-3}$) and stripped ($Z>0.01\,Z_{\odot}$, $n<0.05\,{\rm cm}^{-3}$). Indeed, at $z=3$ and 5 the mass accretion rates in cold streams is very high ($\dot M \simeq 20\,M_\odot$yr$^{-1}$). A significant amount of baryons are also accreted from stripped satellites, although quantifying this amount is difficult in Eulerian schemes since this metal rich material can mix with the other gas phases that have never been part of the satellites. After $z\sim2$, the hot mode of accretion dominates, making large clump formation at large radii only possible through galaxy mergers, c.f. \cite{Barnes92} and E93. At $z\sim 2$, the galaxy has a thin and extended spiral disk component. Although the gas velocity dispersion is still rather high in the disk, the Jeans mass in the spiral arms is on the order of $\sim10^7\,M_\odot$, closer to the largest giant molecular clouds in present day spiral galaxies. The corresponding gas $Q_{\rm g}$-parameter \citep{goldreichlyndenbell65a} is $Q_{\rm g} \simeq 1.5-2$ in the star forming region, indicating that the disk is marginally stable and the galaxy has reached a quiescent phase with no further large clump formation.

\begin{figure}
\center
\psfig{file=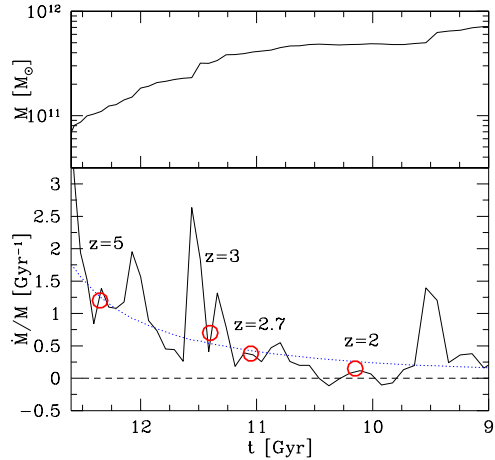,width=185pt}
\caption[]{Top panel: dark matter accretion history [and its logarithmic derivative (bottom panel)]
for the VL-2 halo using the HOP halo finder \citep{EisensteinHut1998}. After a period of moderate mass increase, during several epochs (e.g. $z\sim 3.25$ and $\sim1.75$) the halo mass and hence the gaseous mass dramatically increases. The red rings mark specific times discussed in the text. The dotted blue line shows the expected averaged gas accretion calculated from extended Press-Schechter (EPS) theory \citep{Neistein06}.}
\label{fig:accretion}
\end{figure}

Fig.~\ref{fig:accretion}, shows the dark matter mass accretion rate in the simulated galaxy, as a function of time. At $z=2$, the accretion rate is significantly lower than the average, explaining why the disk has reached this quiescent phase. A global analytical approach for understanding high-$z$ disk fragmentation can be applied \citep{DSC09}, based on simple stability arguments and the disk fraction $\delta\equiv M_{\rm d}/M_{\rm tot}(R_{\rm d})$. Here $M_{\rm d}$ is the baryonic mass in the disk and $M_{\rm tot}(R_{\rm d})$ is the total mass within the disk radius $R_{\rm d}$. A $\delta\sim0.25-0.5$ should give rise to large clumps involving a few percent of the disk mass and $\delta\sim 0.3-0.35$ is predicted for a steady-state fragmentation from moderately clumpy streams. The disk in our simulation at $z=5, 3$ and 2 has $\delta=0.47, 0.33$ and 0.33 respectively which is in excellent agreement with the above prediction (see also Fig. 2 in \cite{DSC09}). Using only the gas in $M_{\rm d}$ renders a lower bound of $\delta=0.17, 0.17$ and 0.1. We point out that the stellar fraction increases significantly towards lower redshifts and this 'hotter' component stabilizes the disk at $z\sim 2$.

\section{Conclusions}
\label{sect:discussion}
We have followed the formation and accretion history of a Milky Way sized galaxy using state-of-the-art AMR techniques. Most of the baryons are in an orderly rotating disk by a redshift $z=2$, but how they attain this equilibrium is very complex and the focus of this work.
One of the most important points of this paper is that we can answer the question in detail of `how galaxies get their baryons'. Extending recent work on the impact of cold streams on galaxy formation \citep{Keres05,DekelBirnboim06,Ocvirk08,Dekel09}, we analyze for the first time how single phase narrow cold streams and ram pressure stripped debris assembles an extended turbulent rotating disk. Complex gas interactions takes place in an extended accretion region in which infalling gas is decelerated through compression/radiative shocks and from the pressure gradients arising from a hot halo component.

Prior to $z\sim2$, the accretion rate of cold gaseous material onto the disk is the highest and
we resolve the gravitational instabilities responsible for the formation of many very massive clumps within an extended $\sim 10\,\kpc$ disk. This is about two times larger than the theoretical expectations of disk sizes at this epoch \citep{MoMaoWhite98}. The observed morphology, star forming rate, global rotation and metallicity of the system is in good agreement with the observed clump-cluster and chain galaxies \citep{Elmegreen06,DebraElmegreen07,Bournaud08}. This scenario is an extension of the disk
fragmentation scenario proposed by B07 and \cite{Elmegreen08}, although here studied more consistently within the current cosmological framework, and more specifically related to cosmological accretion. Therefore, clumpy galaxies should be most frequent at this epoch since massive clump formation stops during the remaining slow accretion phase and the disk evolves quiescently until $z=0$ which will be reported on in a forthcoming paper.

\section*{Acknowledgments}
The authors would like to thank Aaron Boley for valuable discussions on gravitational instability, Fredric Bournaud for discussions on clumpy galaxies and the referee Avishai Dekel for a careful review that improved the quality of the paper. Thanks to J\"urg Diemand for making the initial conditions available and to Doug Potter for making it possible to run the simulations on the zBox supercomputers (http://www.zbox2.org).

\bibliographystyle{mn2e}
\bibliography{chain}
\end{document}